# Searching for Superconductivity in the Z = 4.67 Family

O. P. Isikaku-Ironkwe[1, 2]
[1]The Center for Superconductivity Technologies (TCST)
Department of Physics,
Michael Okpara University of Agriculture, Umudike (MOUAU),
Umuahia, Abia State, Nigeria
and
[2]RTS Technologies, San Diego, CA 92122

## Abstract

Lithium borocarbide, LiBC, a member of the Z = 4.67 family of materials, is not a superconductor. Using ternary combinations of elements that give Z = 4.67 and the material specific characterization dataset (MSCD) scheme, we discover other members of this family. We predict that most of them will not be superconducting and offer an explanation in terms of valence electrons and atomic number ratios. We propose that by appropriately increasing the atomic number while keeping the valence electron count constant, we can transform them into $MgB_2$-like superconductors. We show some examples of such transformations. Alternatively, reducing the valence electron count to 2 while keeping Z constant may lead to high-Tc in the Z=4.67 family.

## Introduction

Magnesium diboride, $MgB_2$, is an unbeaten model of high-Tc binary superconductivity [1, 2]. Many materials exist that have the same structure and valence electrons as $MgB_2$. One of them, LiBC, was predicted to have a Tc very much higher than $MgB_2$ [3] but was found to be not superconducting [4 – 7]. Material specific characterization dataset (MSCD) analysis [9] of LiBC reveals that its average atomic number (Z) is 4.667 and its valence electron count (Ne) is 2.667. In this paper, we search for and examine other materials that have the same average atomic number, Z, and valence electrons, Ne, as LiBC. We show that they too cannot be superconductors. However we show that those with Ne=2.667 can be transformed into superconductors by increasing their average atomic number to 7.333.

## Searching for Z =4.67 Materials

The computation of the MSCD of a material is detailed in [9]. The average $Z = Z_t/A_n$, where $Z_t$ is total atomic number of the elements and $A_n$ is total number of atoms in the formula. A combination of 2 or 3 elements with 3-atoms, whose $Z_t= 14$, produces many chemically stable possibilities. Some of them are LiBC, $BeB_2$, $Be_2C$, $Li_2O$, $MgH_2$ and LiBeN. Moving to 2 or 3 elements with $A_n =6$ and $Z_t= 28$, produces $LiB_5$, $Li_3BN_2$, $NaAlH_4$ and $KBH_4$. We find that the average Z=4.67 family of materials falls into two classes: those with Ne =2.667 and those with Ne= 1.333. The MSCDs of 10 members of this family of materials is displayed in Table 1.

## Superconductivity and Z =4.667 Family

The likelihood of superconductivity in a material [9] may be estimated from knowledge of the Ne and Z of the material by the empirical expression:

$$0.75 < Ne/\sqrt{Z} < 1.02 \qquad (1)$$

We found [9] that most inorganic materials outside this range are not high-Tc superconductors. At the lower end they may be low Tc superconductors under pressure or semiconducting. At the upper range they are not superconducting but may be ferromagnetic. From Table 1 showing the MSCDs of the Z=4.667 family, we see that 7 of them have $Ne/\sqrt{Z}$ =1.2344 which is much higher than the upper range for superconductivity. The other 3 have $Ne/\sqrt{Z}$ =0.6172, which is far below the lower range for superconductivity. We predict that they may not be superconducting under normal conditions.

## Transforming Z=4.67 to Z =7.33

LiBC was predicted to be a higher Tc material than $MgB_2$ [3] based on structural and electronic similarity and its lighter Z. The effort to dope LiBC to produce a higher Tc equivalent in $Li_{0.5}BC$ was not successful [4 – 7]. From our studies, we see from figures 1 to 6 that members of the Z =4.67 can be transformed to Z = 7.33 materials. These new Z=7.33 materials meet the condition for superconductivity of equation (1). Some of them have already been predicted to be MgB$_2$-like superconductors [10 -15].

## Discussion

From the MSCD of the Z=4.667 materials, we observe the close match of the formula weight (Fw) of the first 5 listed materials. They all also have the same electronegativity ($x$=1.8333). We expect them to behave alike [9] as far as superconductivity is concerned. The 6-atom materials have higher Fw and thus higher Fw/Z as is to be expected when number of atoms in a formula increases. The materials with Ne =1.333 show Ne/$\sqrt{Z}$ = 0.6172, still outside the realm of superconductivity. Those materials are known strong oxidizing agents. If we could get materials with Ne=2.0 and Z=4.667, we will get Ne/$\sqrt{Z}$ =0.9258 which is fully within the bounds of superconductivity from equality (1). Combinations of materials within this family may yield stable new superconducting materials with Ne/$\sqrt{Z}$ =0.9258. This will need to be explored further. However the transformations via element substitution and subsequent increase in Z to 7.333 may be the route to getting MgB$_2$-like superconductivity.

## Conclusion

Members of the Z= 4.67 family do not meet the condition for superconductivity that is 0.75<Ne/$\sqrt{Z}$ <1.02. LiBC and 6 members of the Z =4.67 family can be transformed by appropriate element change to Z=7.33 family, without changing the valence electron count. The Z =7.33 has many materials that meet the condition for superconductivity through such transformations.

## Acknowledgements

Many interactions and discussions with M. J. Schaffer, then at General Atomics, M. Brian Maple and J. Hirsch at UC San Diego and J.R. O'Brien at Quantum Design, San Diego and A.O.E. Animalu at University of Nigeria, have been continually useful in the search for novel superconductors.

## Tables

| | Material | $\chi$ | Ne | Z | Ne/$\sqrt{Z}$ | Fw | Fw/Z |
|---|---|---|---|---|---|---|---|
| 1 | LiBC | 1.8333 | 2.6667 | 4.6667 | 1.2344 | 29.76 | 6.377 |
| 2 | BeB$_2$ | 1.8333 | 2.6667 | 4.6667 | 1.2344 | 30.63 | 6.564 |
| 3 | Be$_2$C | 1.8333 | 2.6667 | 4.6667 | 1.2344 | 30.03 | 6.435 |
| 4 | Li$_2$O | 1.8333 | 2.6667 | 4.6667 | 1.2344 | 29.88 | 6.403 |
| 5 | LiBeN | 1.8333 | 2.6667 | 4.6667 | 1.2344 | 29.96 | 6.420 |
| 6 | LiB$_5$ | 1.8333 | 2.6667 | 4.6667 | 1.2344 | 60.99 | 13.07 |
| 7 | Li$_3$BN$_2$ | 1.8333 | 2.6667 | 4.6667 | 1.2344 | 59.65 | 12.78 |
| 8 | NaAlH$_4$ | 1.8 | 1.3333 | 4.6667 | 0.6172 | 54.01 | 11.574 |
| 9 | KBH$_4$ | 1.8667 | 1.3333 | 4.6667 | 0.6172 | 53.95 | 11.56 |
| 10 | MgH$_2$ | 1.8 | 1.3333 | 4.6667 | 0.6172 | 26.33 | 5.642 |

**Table 1:** MSCD of Z =4.667 Materials. $\chi$ is the average electronegativity, Ne is the averages valence electron count, Z the average atomic number and Fw the formula weight of the material. Computation of MSCD is fully described in [9]. Note that seven of Z =4.67 materials have same electronegativity and Ne = 2.667 while the remaining three have Ne =1.333 and varied electronegativity.

| | Material | $\chi$ | Ne | Z | Ne/$\sqrt{Z}$ | Fw | Fw/Z |
|---|---|---|---|---|---|---|---|
| 1 | LiAlC | 1.667 | 2.6667 | 7.333 | 0.9847 | 45.93 | 6.263 |
| 2 | NaBC | 1.8 | 2.6667 | 7.333 | 0.9847 | 45.81 | 6.247 |
| 3 | BeBAl | 1.667 | 2.6667 | 7.333 | 0.9847 | 46.80 | 6.382 |
| 4 | MgB$_2$ | 1.733 | 2.6667 | 7.333 | 0.9847 | 45.93 | 6.263 |
| 5 | Li$_2$S | 1.5 | 2.6667 | 7.333 | 0.9847 | 45.95 | 6.266 |
| 6 | Be$_2$Si | 1.6 | 2.6667 | 7.333 | 0.9847 | 46.11 | 6.290 |
| 7 | LiBeP | 1.533 | 2.6667 | 7.333 | 0.9847 | 46.92 | 6.398 |
| 8 | LiMgN | 1.733 | 2.6667 | 7.333 | 0.9847 | 45.26 | 5.999 |
| 9 | KB$_5$ | 1.8 | 2.6667 | 7.333 | 0.9847 | 93.15 | 12.702 |

**Table 2:** MSCDs of 9 products from the transformations of Z =4.67 non-superconductors to Z =7.33 materials. MgB$_2$ is a known superconductor and the others are predicted superconductors [9 - 15].

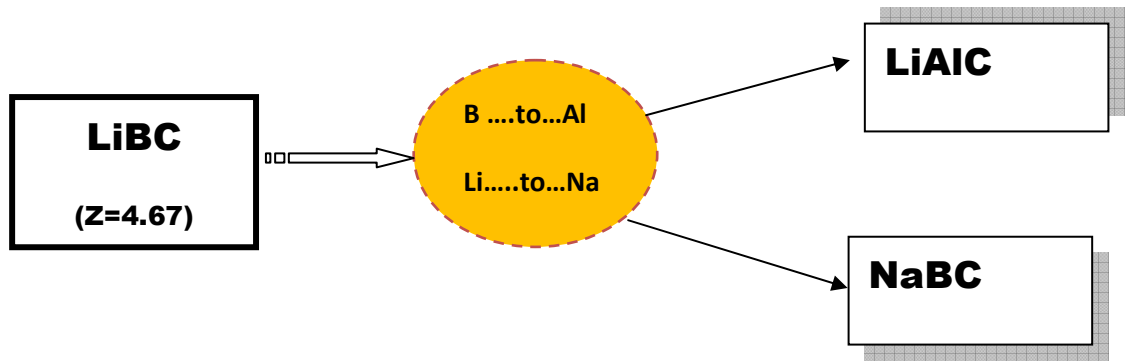

**Figure 1**: LiBC can be transformed to LiAlC by substituting Al for B keeping Ne =2.67. It can also be transformed to NaBC by substituting Na for Li, keeping Ne=2.67. In both cases Z changes from 4.67 to 7.33. The products LiAlC and NaBC should have $Ne/\sqrt{Z} = 0.9847$ which is within the bounds for superconductivity (0.75<$Ne/\sqrt{Z} < 1.02$). See Table 2.

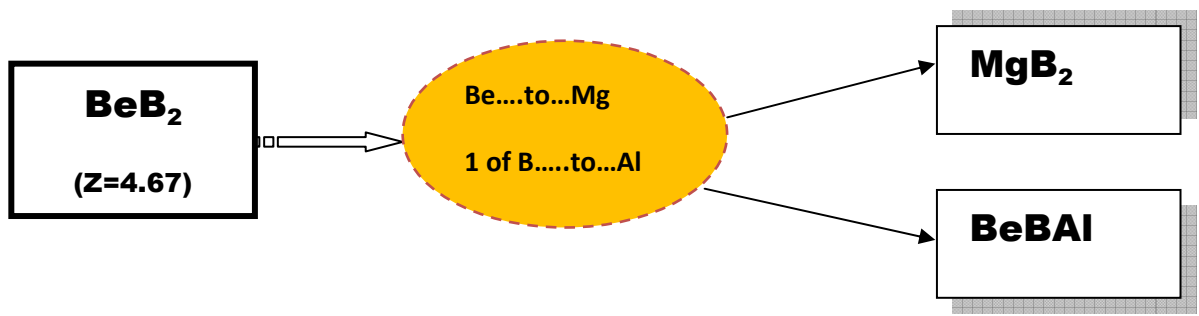

**Figure 2**: $BeB_2$ can be transformed to $MgB_2$ by substituting Mg for Be keeping Ne =2.67. It can also be transformed to BeBAl by substituting Al for one B, keeping Ne=2.67. In both cases Z changes from 4.67 to 7.33. The products $MgB_2$ and BeBAl should have $Ne/\sqrt{Z} = 0.9847$ which is within the bounds for superconductivity. See Table 2.

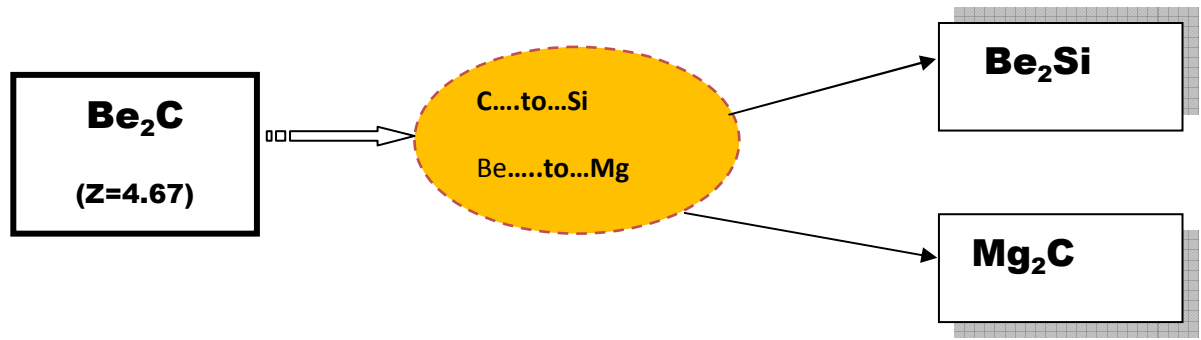

**Figure 3**: Be$_2$C can be transformed to Be$_2$Si by substituting Si for C keeping Ne =2.67. It can also be transformed to Mg$_2$C by substituting Mg for Be, keeping Ne=2.67. In one case Z changes from 4.67 to 7.33. In the other case, Z changes from 4.67 to 10. The products Be$_2$Si and Mg$_2$C should have 0.75<Ne/$\sqrt{Z} < 1.02$ which is within the bounds for superconductivity. See Table 2.

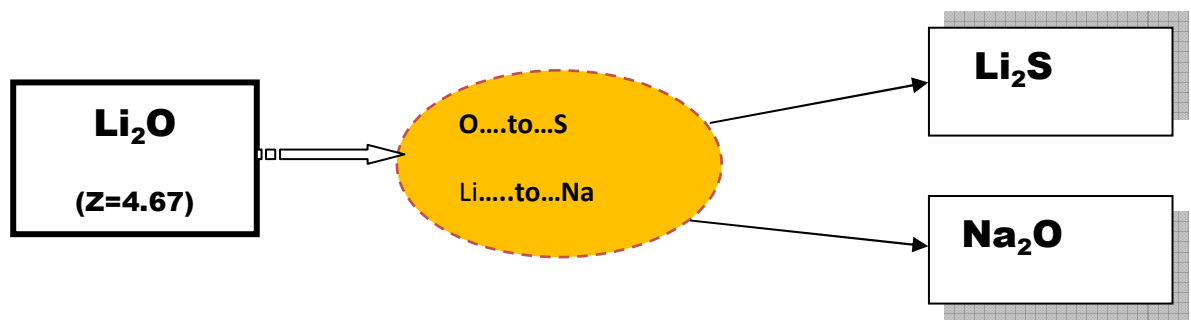

**Figure 4**: Li$_2$O can be transformed to Li$_2$S by substituting S for O keeping Ne =2.67. It can also be transformed to Na$_2$O by substituting Na for Li, keeping Ne=2.67. In one case Z changes from 4.67 to 7.33. In the second case, Z changes from 4.67 to 10. The products Li$_2$S and Na$_2$O should have 0.75<Ne/$\sqrt{Z} < 1.02$ which is within the bounds for superconductivity. See Table 2.

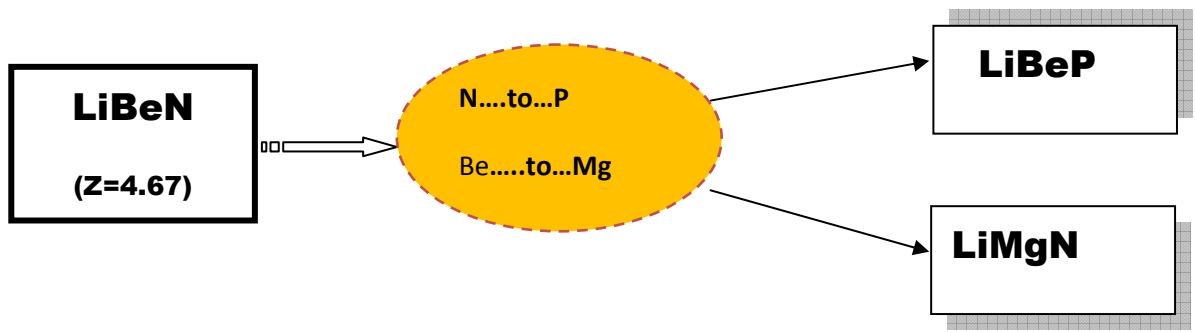

**Figure 5**: LiBeN can be transformed to LiBeP by substituting P for N keeping Ne =2.67  In this case Z changes from 4.67 to 7.33. Also by substituting Mg for Be we get LiMgN which has Z =7.333. The products LiBeP and LiMgN have 0.75<Ne/$\sqrt{Z}$ < 1.02 which is within the bounds for superconductivity. See Table 2.

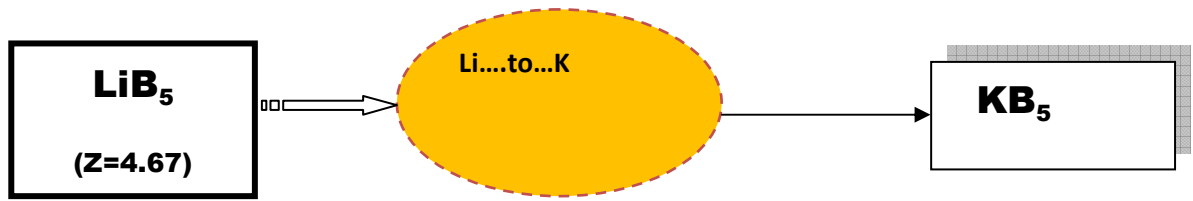

**Figure 6**: LiB$_5$ can be transformed to KB$_5$ by substituting K for Li keeping Ne =2.67  In this case Z changes from 4.67 to 7.33. The product KB$_5$ has 0.75<Ne/$\sqrt{Z}$ < 1.02 which is within the bounds for superconductivity. See Table 2.